\documentclass[12pt]{article}
\usepackage{amsmath}
\usepackage{amssymb}

\newcommand{\be}{\begin{equation}}
\newcommand{\bea}{\begin{eqnarray}}
\newcommand{\ee}{\end{equation}}
\newcommand{\eea}{\end{eqnarray}}
\def\theequation{\arabic{section}.\arabic{equation}}
\newcommand{\hp}{\hat{p}}

\newcommand{\he}{\hat{\epsilon}}
\newcommand{\apr}{\alpha'}
\newcommand{\bn}[2]{\left(\begin{array}{c} #1\\ #2
\end{array}\right)}

\begin{document}
\topmargin -1cm \oddsidemargin=0.25cm\evensidemargin=0.25cm
\setcounter{page}0
\renewcommand{\thefootnote}{\fnsymbol{footnote}}
\begin{titlepage}
\begin{flushright}
DFTT-12/2010\\
LTH-886
\end{flushright}
\vskip .7in
\begin{center}
{\Large \bf On the Tensionless Limit of String theory, Off --
Shell Higher Spin Interaction Vertices and BCFW Recursion
Relations } \vskip .7in {\large Angelos
Fotopoulos$^a$\footnote{e-mail: {\tt foto@to.infn.it}} and Mirian
Tsulaia$^{b,c}$\footnote{e-mail: {\tt tsulaia@liv.ac.uk}, Associate
member of the Centre of Particle Physics and Cosmology, Ilia State
University, 0162, Tbilisi, Georgia }} \vskip .2in {$^a$ \it
Dipartimento di Fisica Teorica dell'Universit\`a di Torino and
INFN Sezione di Torino via
P.Giuria 1, I-10125 Torino, Italy} \\
\vskip .2in { $^b$ \it Department of Mathematical Sciences, University of Liverpool,
Liverpool, L69 7ZL, United Kingdom }\\
\vskip .2in { $^c$ \it Centre for Theoretical Chemistry and Physics, Massey University of Auckland,
Private Bag 102904, 0745 Auckland, New Zealand}
\\

\begin{abstract}
 We construct an off--shell extension of cubic interaction  vertices between massless
bosonic Higher Spin fields on a flat background which can be
obtained from perturbative bosonic string theory. We demonstrate
how to construct higher quartic interaction vertices using a
simple particular example. We examine whether BCFW recursion
relations for interacting Higher Spin theories are applicable. We
argue that for several interesting examples such relations should
exist, but consistency of the theories might require that we
supplement Higher Spin field theories with extended and possibly
non-local objects.

\end{abstract}

\end{center}

\vfill

\end{titlepage}

\tableofcontents

\section{Introduction}

Higher Spin (HS) gauge theories (see \cite{Vasiliev:2004qz}--\cite{Fotopoulos:2008ka} for recent reviews)
 are usually formulated in two different ways: in frame--like \cite{Fradkin:1986qy} -- \cite{Bandos:1998vz}
or in  metric--like \cite{Fronsdal:1978rb} -- \cite{Francia:2002pt} approaches.
Although a formulation of free dynamics of various representations of Poincare and anti de Sitter
groups in both approaches is quite a  nontrivial task, the most challenging problem is the construction and study of interactions
between both massless and massive Higher Spin fields on flat and curved backgrounds.

A landmark has been reached in \cite{Fradkin:1986qy} with the understanding that, the anti de Sitter background can accommodate consistent
self interactions of massless Higher Spin fields. The interaction involves an infinite
tower of massless Higher Spin fields and is non local. Moreover the presence of an anti de Sitter background, where no
S-matrix can be defined, allows to naturally bypass Coleman--Mandula no--go theorem. These results, apart form being remarkable in their own right,
can  also be extremely useful for a better understanding of String/M theory in particular in the context of AdS/CFT correspondence
\cite{Sezgin:2002rt}--\cite{Giombi:2009wh}.

The interaction of massless Higher Spin fields on a flat
background has  always been considered to be more problematic than
the interaction on AdS space. However, this problem  has  recently
been extensively discussed  in
\cite{Metsaev:2007rn}--\cite{Polyakov:2010qs} and several
interesting cubic vertices have been obtained. These results
indicate that one can possibly have consistent interacting
theories of massless Higher Spin fields on a Minkowski background
as well, provided that, as in the case of an AdS background, one
has an infinite number of  fields and the interactions are non
local.

It is extremely important to find a guiding principle to construct
interactions on flat space time. A possible approach is to
consider perturbative String Theory as a ``laboratory'' for
studying interactions between massless Higher Spin fields
\cite{Sagnotti:2010at}--\cite{Polyakov:2010qs},
\cite{Fotopoulos:2007nm}. Obviously since in the usual tensile
string theory all Higher Spin fields are massive,  one needs to go
to a limit where the masses of Higher Spin fields go to zero.
However the high energy limit of string theory
\cite{Gross:1987ar}-- \cite{Bonelli:2003kh} is still to be better
understood\footnote{Related work on the tensionless limit of
string theory has been done in \cite{Lindstrom:2003mg}}.

In  a recent paper \cite{Sagnotti:2010at}, the authors constructed
cubic vertices for HS fields via the high energy limit of string
perturbation theory. Their result has as a part of it the vertex
constructed in \cite{Fotopoulos:2007nm} via the high energy limit
of Open String Field Theory (OSFT), but includes additional terms
which grant the new vertex with a non-abelian structure. These
results are extremely important to further study the problem of
interactions between massless fields on a flat background, since
the well developed technique of string perturbation theory can
allow the study of higher order interactions and perform
nontrivial consistency checks of the theory. One of the aims of
the present paper is to demonstrate how one can construct vertices
which describe a consistent cubic interaction of massless bosonic
Higher Spin fields, serving as an off --shell completion of the on
--shell vertex derived in \cite{Sagnotti:2010at}. We are using the
BRST method  for interacting higher spin fields, which can be
formulated in flat and in AdS spaces \cite{Bengtsson:1987jt},
\cite{Buchbinder:2006eq} -- \cite{Fotopoulos:2009iw}. This method
is essentially based on the requirement of  gauge invariance and
the abelian gauge transformations of the free theory are deformed
to nonlinear (generically nonabelian) transformations to obtain
nontrivial cubic and higher order interactions.

Having formulated an off--shell description for a system which
describes cubic interactions between massless Higher Spin fields
 belonging to reducible representations of the Poincare group,
one can further consider higher order interactions for massless
Higher Spin Fields following standard perturbation theory. This
consideration will hopefully improve our understanding of the
problem of interactions in Higher Spin theories on Minkowski
background, since this  approach can be further generalized to
higher (hopefully arbitrary) orders in the coupling constant.
Therefore we consider the present article as  a step towards a
better understanding and towards further consistency checks of the
interacting theory of massless Higher Spin fields on a flat
background. As a particular example we show how to construct a
quartic interaction vertex from a cubic Lagrangian described in
\cite{Koh:1986vg}. This problem,  although being technically
simpler than the problem of construction of quartic interaction
vertices from  cubic ones of the type \cite{Sagnotti:2010at}, is
still quite instructive and contains some lessons on how to deal
with more complicated systems.

In the second part of the paper we will follow a different
approach to the problem based on the S--matrix rather than the
Lagrangian itself. Recently, there has been remarkable progress in
exploring the properties of the S--matrix for tree level
scattering amplitudes. Motivated by Witten's twistor formulation
of ${\cal N}=4$ Super--Yang--Mills (SYM) \cite{Witten:2003nn},
several new methods have emerged which allow one to compute tree
level scattering amplitudes for gauge and gravity theories. In
particular, the CSW method of \cite{Cachazo:2004kj} has
demonstrated how one can use the Maximum Helicity Violating (MHV)
amplitudes of \cite{Parke:1986gb} as field theory vertices to
construct arbitrary gluon amplitudes.

Analyticity of gauge theory tree level amplitudes has led to the
BCFW recursion relations \cite{Britto:2004ap, Britto:2005fq,
Cachazo:2005ga}. Specifically, analytic continuation of external
momenta in a scattering amplitude allows one, under certain
assumptions, to determine the amplitude through its residues on
the complex plane. Locality and unitarity require that the residue
at the poles is a product of lower-point on shell amplitudes.
Actually the  CSW construction turns out to be a particular
application of the BCFW method \cite{Risager:2005vk}.

The recursion relations of \cite{Britto:2005fq} are at the heart
of many of the aforementioned developments. Nevertheless, for these relations
 the asymptotic behavior of amplitudes under
complex deformation of some external momenta is crucial. When the
complex parameter, which parametrizes the deformation, is taken to
infinity an amplitude should fall sufficiently fast so that a pole
at infinity will be absent\footnote{There has been though some
recent progress  \cite{Feng:2009ei,Feng:2010ku}
 in generalizing the  BCFW relations for theories with boundary
contributions.}. Although naive power counting of individual
Feynman diagrams seems to lead to badly divergent amplitudes for
large complex momenta, it is intricate cancellations among them
which lead to a much softer behavior than expected. In particular, gauge
invariance and supersymmetry in some cases are responsible for
these cancellations.

It is natural to wonder whether these field theoretic methods can
be applied and shed some light into the structure of string theory
amplitudes. This is motivated, in particular, by the fact that the
${\cal N}=4$ SYM theory that plays a central role in the
developments we described above, appears as the low-energy limit
of  string theory in the presence of  D3-branes. In order to even
consider applying the aforementioned methods to string scattering
amplitudes, one needs as a first step to study the large complex
momentum behavior of these amplitudes. Since generally string
amplitudes are known to have excellent large momentum behavior,
one expects that recursion relations should be applicable here as
well. Nevertheless, one should keep in mind that although the
asymptotic amplitude behavior might be better than any local field
theory, the actual recursion relations are quite more involved.
They require knowledge of an infinite set of on-shell string
amplitudes, at least the three point functions, between arbitrary
Regge trajectory states of string theory.

The study of the asymptotic behavior of string amplitudes under
complex momentum deformations was initiated in \cite{Boels:2008fc}
and further on elaborated in \cite{Cheung:2010vn,Boels:2010bv}. In
these works it was established, using  direct study of the
amplitudes in parallel with  Pomeron techniques, that both open and
closed bosonic and supersymmetric string theories have good
behavior asymptotically, therefore  allowing one to use the BCFW
method. So, in  light of the apparent relation of massless Higher
Spin theories with the high energy limit of string theory one
might be able to use these powerful tools in order to study the
properties of HS theories themselves. We will argue that indeed
the arguments which suggest that the boundary contribution under
BCFW deformation of string amplitudes vanishes, extend to the
corresponding Higher Spin amplitudes. The failure though of the
criterion of \cite{Benincasa:2007xk} will indicate that the Higher
Spin theory on a flat background  might be missing crucial
ingredients required for its consistency at the full interaction
level. In two examples we will derive some indications that
consistent HS theories cannot rely solely on massless HS point
particles in their spectrum but might require extra degrees of
freedom like extended objects and/or non-local states with Pomeron
like dynamics.

The paper is organized as follows:

In Section \ref{BE} we collect some of the main concepts of the
BRST constructions for interacting massless Higher Spin fields. In
Section \ref{Cubic} we show, using the results of
\cite{Bengtsson:1987jt},
 how to construct a cubic off -- shell interaction vertex which corresponds
to the --on shell vertex \cite{Sagnotti:2010at}, derived from the
string perturbation theory. We also derive a possible
generalization of this vertex to a system which contains bosonic
massless fields which belong to reducible representations of the
Poincare group with mixed symmetry. In Section \ref{QUAR} we
derive a quartic interaction vertex for the cubic Lagrangian
constructed in \cite{Koh:1986vg}. We comment that also in this
case, like the exact cubic vertex of \cite{Fotopoulos:2007nm}, the
algebra is abelian and it leads to a trivial interaction beyond
the cubic level. In the last two sections we will give a short
review of the BCFW method for field theories and we will apply
this method on two different cubic couplings which will allow us
to make some interesting observations on the possible theories
with interacting HS particles on the flat background.
 We conclude with a summary of our
results and some comments.

\setcounter{equation}0\section{Basic equations}\label{BE} Let us
 start with the BRST charge for the open bosonic string
(see for example \cite{Fotopoulos:2008ka} for more details)
\begin{equation}\label{B1}
{\cal Q} = \sum_{k, l=-\infty}^{+\infty}(C_{-k} L_k - \frac{1}{2}
(k-l):C_{-k}C_{-l}B_{k+l}:)-C_0,
\end{equation}
 perform the rescaling of  oscillator variables
\begin{equation}\label{B2}
c_k = \sqrt{2 \alpha^\prime}C_k, \qquad b_k = \frac{1}{\sqrt{2
\alpha^\prime}}B_k, \qquad c_0 =  \alpha^\prime C_0, \qquad b_0 =
\frac{1}{ \alpha^\prime}B_0,
\end{equation}
$$
\alpha^\mu_k \rightarrow  \sqrt{k} \alpha^\mu_k
$$
and then take a formal limit $\alpha^\prime \rightarrow \infty$. In this way one
obtains a BRST charge
\begin{equation}\label{hebrst}
 Q = c_0 l_0 + \tilde{Q} - b_0 {\cal M}
\end{equation}
\begin{equation}\label{B3}
\tilde{Q} = \sum_{k=1}^\infty ( c_k l^{+}_k + c_k^+ l_{k} ), \quad
{\cal M}=\sum_{k=1}^\infty c^{+}_k c_k, \quad l_0 = p^\mu p_\mu,
\qquad l_k^{+}= p^\mu \alpha_{k \mu}^+
\end{equation}
which is nilpotent in any space-time dimension. The oscillator
variables obey the usual (anti)commutator relations
\begin{equation}\label{B4}
[\alpha_\mu^k, \alpha_\nu^{l,+} ] = \delta^{kl} \eta_{\mu \nu},
\quad \{ c^{k,+}, b^l \} = \{ c^k, b^{l,+} \} = \{ c_0^k , b_0^l \}
= \delta^{kl}\,,
\end{equation}
and the vacuum in the Hilbert space is defined as
\begin{equation}\label{B5}
\alpha^\mu_k |0\rangle =  0, \quad
 c_k|0\rangle =  0  \quad k>0 , \qquad
b_k|0\rangle\ = \ 0 \qquad k \geq 0.
\end{equation}
Let us note that one can take the value of $k$ to be any fixed
number without affecting the nilpotency of the BRST charge
(\ref{hebrst}). Fixing the value $k=1$ one obtains the description
of totally symmetric massless higher spin fields, with spins $s,
s-2,..1/0$. The string functional (named ''triplet"
\cite{Francia:2002pt}) in this simplest case has the form
\begin{equation}
\label{Phifield} |\Phi  \rangle = |\phi_1\rangle + c_0
|\phi_2\rangle=  |\varphi \rangle + c^+ \ b^+\ |d\rangle + c_0\
b^+\ |c\rangle
\end{equation}
whereas for an arbitrary value of $k$ one has the so called
''generalized triplet"
\begin{equation}
\label{gentri} |\Phi  \rangle =
 \frac{c^+_{k_1}\dots c^+_{k_p} b^+_{l_1}\dots b^+_{l_p}}{{(p!)}^2}
|D^{l_1, \dots l_p}_{k_1, \dots l_p}\rangle + \frac{c_0
c^+_{k_1}\dots c^+_{k_{p-1}} b^+_{l_1}\dots b^+_{l_p}}{(p-1)! p!}
|C^{l_1, \dots l_p}_{k_1, \dots k_{p-1}}\rangle,
\end{equation}
where the vectors $|D^{k_1, \dots k_p}_{l_1, \dots l_p}\rangle $ and
$|C^{k_1, \dots k_p}_{l_1, \dots l_p}\rangle $ are expanded only in
terms of oscillators $\alpha^{\mu +}_k$, and the first term in the
ghost expansion of (\ref{gentri}) with $p=0$ corresponds to the
state $|\varphi \rangle$ in (\ref{Phifield}). One can show that
 the whole spectrum of the open bosonic string decomposes
into an infinite number of generalized triplets, each of them
 describing a finite number of fields with mixed symmetries
\cite{Sagnotti:2003qa}.

In order to describe the cubic interactions  one introduces
 three copies ($i=1,2,3$) of the Hilbert space
defined above, as in bosonic OSFT \cite{Gross:1986ia}. Then the
Lagrangian has the form
\begin{equation} \label {LIBRST}
{L} \ = \ \sum_{i=1}^3 \int d c_0^i \langle \Phi_i |\, Q_i \,
|\Phi_i \rangle \ + g( \int dc_0^1 dc_0^2  dc_0^3 \langle \Phi_1|
\langle \Phi_2|\langle \Phi_3||V_3 \rangle + h.c)\,, \end{equation}
where $|V_3\rangle$ is the cubic vertex and $g$ is a
 coupling
constant. The Lagrangian (\ref{LIBRST}) is  invariant at order $g$
with respect to the nonabelian gauge transformations
\begin{equation}\label{BRSTIGT1}
\delta | \Phi_i \rangle  =  Q_i | \Lambda_i \rangle  - g \int
dc_0^{i+1} dc_0^{i+2}[(  \langle \Phi_{i+1}|\langle \Lambda_{i+2}|
+\langle \Phi_{i+2}|\langle \Lambda_{i+1}|) |V_3 \rangle] \,,
\end{equation}
provided that the vertex $|V\rangle$ satisfies the BRST invariance
condition
\begin{equation}\label{VBRST}
\sum_i Q_i |V_3 \rangle=0\,.
\end{equation}
 The
gauge parameter $|\Lambda \rangle$ in each individual Hilbert space
has the ghost structure
\begin{equation}\label{GPdef}
|\Lambda \rangle = b^+ |\lambda\rangle
\end{equation}
 for the totally symmetric case, while the gauge parameters for the generalized triplets
 take the form
\begin{equation}
\label{genGPdef} |\Lambda  \rangle =
 \frac{c^+_{k_1}\dots c^+_{k_p} b^+_{l_1}\dots b^+_{l_{p+1}}}{(p!)(p+1)!}
|\Lambda^{l_1, \dots l_{p+1}}_{k_1, \dots k_p}\rangle + \frac{c_0
c^+_{k_1}\dots c^+_{k_{p-1}} b^+_{l_1}\dots c^+_{l_{p+1}}}{(p-1)!
(p+1)!} |\hat{\Lambda}^{l_1, \dots l_{p+1}}_{k_1, \dots
k_{p-1}}\rangle. \nonumber
\end{equation}

Let us make some comments about the BRST charge (\ref{hebrst}). We
can actually justify the way it was obtained from the BRST charge of
the open bosonic string since its cohomologies correctly describe
equations of motion for massless bosonic fields belonging to mixed
symmetry representations of the Poincare group (see e.g.
\cite{Sagnotti:2003qa}). So taking the point of view that, in the
high energy limit the whole spectrum of the bosonic string collapses
to zero mass, which is now infinitely degenerate, one can take the
BRST charge (\ref{hebrst}) as the one which correctly describes this
spectrum.

\setcounter{equation}0\section{Cubic vertex for HS interactions
from the high energy limit of String Theory}\label{Cubic}

In this section we will demonstrate how one can construct an
off--shell extension of the cubic vertex suggested in
\cite{Sagnotti:2010at}. The key point is that the free spectrum of
string theory is given by the free Lagrangian of triplets in
(\ref{LIBRST}). This suggests that we use the on--shell vertex of
\cite{Sagnotti:2010at} as an ansatz in the BRST method of the
previous  section for constructing interacting Lagrangians for
triplets. The cubic vertex of \cite{Sagnotti:2010at} is an
extension of the unmodified cubic vertex of \cite{Koh:1986vg,
Fotopoulos:2007nm}.

Let us describe first the proposed high energy limit of string
theory of \cite{Fotopoulos:2007nm}. The cubic vertex in this case
takes the form
\begin{equation}\label{KOansatz}
V^{1}= exp\ (\ Y_{ij} \alpha^{i+}_\mu p_\mu^j + Z_{ij}
c^{i+}b_0^j\ )\,.
\end{equation}
The BRST invariance condition of the vertex  (\ref{KOansatz}) imposes the following
condition on the coefficients $Y_{ij}$ and $Z_{ij}$
\begin{equation}\label{KOsolution}
Z_{i,i+1}+Z_{i,i+2}=0
\end{equation}
$$
Y_{i,i+1}= Y_{ii}-Z_{ii} -1/2(Z_{i,i+1}-Z_{i,i+2})
$$
$$
Y_{i,i+2}= Y_{ii}-Z_{ii} +1/2(Z_{i,i+1}-Z_{i,i+2}).
$$

There are two options available at this point. One is to consider
the modification of the cubic Lagrangian (\ref{LIBRST}) and gauge
transformations (\ref{BRSTIGT1}) which involve quartic and higher
order vertices. We shall consider this option in the chapter 4.
Another option is  to make this solution exact in all orders of
the  coupling constant $g$. For this we have to modify the
solution above by an extra oscillator-- dependent factor i.e.,
consider a modified ansatz \cite{Fotopoulos:2007nm}
\begin{equation}\label{KOmod}
|V \rangle=  V^{1} \times V^{mod}c_0^1 c_0^2 c_0^3|0\rangle_{123}
\end{equation}
with
\begin{equation}\label{Vmod}
V^{mod}=exp\ (\ S_{ij} c^{i+} b^{j+} + \frac{P_{ij}}{2} \alpha^{i+}_\mu \alpha^{j+}_\mu\ ).
\end{equation}
The BRST invariance condition imposes the constraint on the coefficients
$S_{ij}$ and $P_{ij}$
\begin{eqnarray}\label{KOmsol}
&& S_{ij}= P_{ij}=0 \qquad i\neq j \\
&& P_{ii} - S_{ii}=0 \qquad i=1,2,3 \nonumber
\end{eqnarray}
whereas the requirement that the solution (\ref{Vmod}) is exact in
all orders in the coupling constant further restricts the value of
the parameter $S_{ii}$ to be equal to $1$.

 This vertex leads into abelian, but
nonlinear gauge transformations. Therefore, one can verify using
the technology developed in \cite{Fotopoulos:2009iw}, that the
four point function of such a theory turns out to be vanishing and
makes the theory not a good framework for the study of non-trivial
interacting HS theories at quartic and higher levels. In the next
section we will try to relax the condition of an exact cubic
vertex and allow a quartic vertex in our Lagrangian and possibly
higher ones in an attempt to derive a non-trivial algebra and
therefore non-trivial interactions.

In an alternative method given in \cite{Sagnotti:2010at} one starts
from the on--shell string amplitudes for the leading Regge
trajectory of string theory and employs the tensionless limit
through  the study of decoupling of null states. An
on--shell vertex obtained using this method  has a leading term which is given by going
on--shell in (\ref{KOansatz}). The  derived subleading terms
involve an exponential of higher terms in the oscillators and lead
in a non--trivial algebra of gauge transformations. In the
following subsection \ref{KOm} we will derive an off--shell
expression of the vertex given in \cite{Sagnotti:2010at} for the
leading Regge trajectory and in subsection \ref{KOmixed} we will
show how one can derive the cubic vertex for mixed symmetry HS
fields. This  result is an ansatz for the on--shell vertex for
subleading Regge trajectories in the spirit of
\cite{Sagnotti:2010at}.

\subsection{A cubic vertex for totally
symmetric fields}\label{KOm}

Since the ``leading''  vertex (\ref{KOansatz}) is BRST invariant on its
own right we will consider the BRST invariance of the subleading
modification suggested in \cite{Sagnotti:2010at}.  We begin first
with the simpler case of a vertex for totally symmetric fields.
This means that we consider only one set of oscillators as in
(\ref{B4}). Let us note that one can make the functionals $|
\Phi_i \rangle$ matrix valued i.e., add Chan--Paton factors, but we
shall not do so.

In order to make a connection with the on -- shell vertex of
\cite{Sagnotti:2010at} let us make the following ansatz
\begin{equation}\label{Bansatz1}
|V_3 \rangle = e^{v} c_0^1 c_0^2 c_0^3 | 0 \rangle_{123}\,.
\end{equation}
where, (we follow the conventions of \cite{Buchbinder:2006eq}), \bea
\label{Bansatz2} \nonumber v& =& X^{(1)}_{rstu} (\alpha^{r+}_\mu
\alpha^{s+}_\mu) (\alpha^{t+}_\nu p_{\nu}^u)  +
X^{(2)}_{rstu} (c^{r+} b^{s+}) (\alpha^{t+}_\mu p_{\mu}^u) + \\
&&X^{(3)}_{rstu} (\alpha^{r+}_\mu \alpha^{s+}_\mu) (c^{t+} b_0^u)
+ X^{(5)}_{rstu} (c^{r+} b^{s+}) (c^{t+} b_0^u) \eea The
terms in the expression above proportional to $X^{(1)}_{rstu}$
correspond to the operator ${\cal G}$ in \cite{Sagnotti:2010at}.
The remaining terms are those required for extending off-shell
their cubic on-shell description.

The coefficients $X^{(1)}_{rstu}, X^{(2)}_{rstu}, X^{(3)}_{rstu}$
and $X^{(5)}_{rstu}$ obey symmetry relations \be \label{SIMS}
X^{(1)}_{rstu}= X^{(1)}_{srtu}, \quad X^{(3)}_{rstu}=
X^{(3)}_{srtu}, \quad  X^{(5)}_{rstu}= - X^{(5)}_{tsru} \ . \ee
They are constants to be determined form the BRST invariance
condition of the vertex. The BRST invariance condition of the
vertex implies \be (2 X^{(1)}_{rstu} p_\mu^r p_\nu^u -
X^{(2)}_{rstu} p_\mu^s p_\nu^u ) c^{r+} \alpha^{s+}_\mu
\alpha^{t+}_\nu=0, \ee \be (- X^{(3)}_{rstu} p_\mu^u p_\mu^u +
X^{(1)}_{rstu} p_\mu^t p_\mu^u ) c^{t+} \alpha^{r+}_\nu
\alpha^{s+}_\nu=0, \ee \be (- X^{(5)}_{rstu} p_\mu^u p_\mu^u +
X^{(2)}_{rstu} p_\mu^t p_\mu^u ) c^{r+} b^{s+}c^{t+}=0, \ee \be (-
X^{(2)}_{rtsu} b_0^t p_\mu^u +2 X^{(3)}_{rstu} b_0^u   p_\mu^r -
X^{(5)}_{rstu} b_0^u   p_\mu^s ) c^{r+} c^{t+} \alpha^{s+}_\mu =0,
\ee \be
 X^{(5)}_{rstu}b_0^s c^{s+} c^{r+} c^{t+}b_0^u=0.
\ee After imposing the cyclic symmetry on the coefficients
$X^{(1)}_{rstu}$ under the indexes $r,s,t,u$ these equations  can
be solved  \cite{Bengtsson:1987jt} to give a solution given in the
Table 1. We can use the method of \cite{Fotopoulos:2009iw} to
decompose our interacting Lagrangian in terms of irreducible
Fronsdal modes. We will get a series of interacting Lagrangians
for each mode which serve as a off-shell extension of the cubic
vertex in \cite{Sagnotti:2010at}. Alternatively if we integrate
out the auxiliary fields $C$ and go on shell for the triplet
fields as in \cite{Francia:2002pt}, we can derive from
(\ref{Bansatz1}) an on-shell cubic vertex. The vertex for the
highest spin mode from each triplet should be the same with the
corresponding vertex for the irreducible HS fields of
\cite{Sagnotti:2010at}.
\begin{table}
\begin{center}
\begin{tabular}[t]{|c|c|c|c|c|}
Index combination & $X^{(1)}$ & $X^{(2)}$ & $X^{(3)}$ & $X^{(5)}$ \\ 
1231 & 1 & -2 & -1 &  1 \\ 
1232 & -1 & 0 & 1 & -1 \\ 
1233 & 0 & 0 & 0 &  1 \\ 
1211 & 0 & 0 & 1 &  0 \\ 
1212 & -1 & 2 & 0 &  0 \\ 
1213 & -1 & 2 & 0 &  0 \\ 
1221 & 1 &-2 & 0 &  -4 \\ 
1222 & 0 &0 & -1 &  1 \\ 
1223 & 1 &0 & 0 & -1 \\ 
1111 & 0 &0 & 0 & 0 \\ 
1112 & -1 &-2 & 1 & 0 \\ 
1113 & 1 &2 & -1 & 0 \\ 
1121 & -2 &-4 & 1 & 1 \\ 
1122 & -6 &-12 & -5 & -4 \\ 
1123 & 0 &-2 & -1 & -1 \\ 
1131 & 2 &4 & -1 & -1 \\ 
1132 & 0 &2 & 1 & 1 \\ 
1133 & 6 &12 & 5 & 4 \\ 
2131 &  & 0 &  &  \\ 
2132 &  & 2 &  &  \\ 
2133 &  & 0 &  &  \\ 
2111 &  & 0 &  &  \\ 
2112 &  & 2 &  &  \\ 
2113 &  & 0 &  &  \\ 
2121 &  & -2 &  &  \\ 
2122 &  & 0 &  &  \\ 
2123 &  & -2 &  &  \\ 
\end{tabular}
\end{center}
\caption{Empty entries in the table mean that the corresponding
value of the coefficient can be recovered from the ones given in
the table using the cyclic property of indices (for example
$X^{(i)}_{1231}= X^{(i)}_{2312}=X^{(i)}_{3123}$) and symmetry
properties (\ref{SIMS})} \label{tabSpectra}
\end{table}

\subsection{A cubic vertex for mixed symmetry
fields}\label{KOmixed}

It is straightforward to generalize the results of the previous
Section to the case of ``generalized triplets'' i.e., for
reducible representations of the Poincare group with mixed
symmetry.

To this end one can replace indexes $r,s,t$ with  double indexes
${\hat r},{\hat s},{\hat t} $, where each index, say ${\hat r}$
has two values ${\hat r}=(r, \tilde r)$. The index $r$ again numerates three Hilbert spaces
$r=1,2,3$ while the index $\tilde r$ numerates different types of oscillators
$\tilde r = 1,.., \infty$.
The oscillators $\alpha^{r}_\mu$, $c^{r}$ and $b^r$ are replaced
with   $\alpha^{{\hat r}}_\mu$, $c^{{\hat r}}$ and $b^{{\hat r}}$.
However, since one has only three different momenta $p_\mu^r$ and only three different
ghost zero modes $c^r_0$ and $b_0^r$, one has $p_\mu^{{\hat 1}} = p_\mu^{1, \tilde 1}=p_\mu^{1, \tilde 2}= ...p_\mu^{1,\tilde n}$,
where $\tilde n$ is an arbitrary natural number.
The BRST charge now reads as:
\begin{equation}
Q = c_0^i l_0^i + c^{\hat i +}l^{\hat i} +c^{\hat i }l^{\hat i +} - c^{\hat i+} c^{\hat i}b_0^{i}, \quad
l^{ {\hat i} \pm}= \alpha^{{\hat i}\pm}_\mu p_\mu^{\hat i}.
\end{equation}
The symmetry properties of the coefficients $X$  are again the
same \be \label{SIMSM} X^{(1)}_{{\hat r}{\hat s} {\hat t} {u}}=
X^{(1)}_{{\hat s} {\hat r} {\hat t} { u}}, \quad X^{(3)}_{ {\hat
r} {\hat s} {\hat t} {u}}= X^{(3)}_{ {\hat s} {\hat r} {\hat t}
{u}}, \quad X^{(5)}_{ {\hat r} {\hat s} {\hat t} { u}}=- X^{(5)}_{
{\hat t} {\hat s} {\hat r} { u}}. \ee The coefficients
$X^{(1)},X^{(2)}, X^{(3)}, X^{(5)}$ for a vertex, which describe
the interaction between generalized triplets
 can be obtained from  Table 1. For example,
the solution $X^{(1)}_{1231}$ corresponds to $X^{(1)}_{{\hat 1}, {\hat 2}, {\hat 3}, {1}}=
X^{(1)}_{ 1 \tilde r,  2 \tilde s,  3 \tilde t, 1 }$  for $\tilde r, \tilde s, \tilde t,  = 1,..., \tilde n$
etc.

\setcounter{equation}0\ \section{A quartic vertex}\label{QUAR}

Let us consider the simple example  given
 in (\ref{KOansatz}) with the modification (\ref{KOmod}). For simplicity we consider the case for totally
symmetric fields.
 As we mentioned before, although the gauge transformations are non--linear
modifications of the free ones and the vertex cannot be removed via field redefinitions,
 it turns out that the algebra is abelian. One can
verify using the formulas derived in \cite{Fotopoulos:2009iw} that
the four--point function of any HS fields vanishes.

Therefore one can investigate  another option, in particular
instead of modifying the cubic vertex (\ref{KOansatz}) using
(\ref{KOmod}), we can add a quartic vertex to the Lagrangian and
make it gauge invariant up to terms of order $g^2$. One can
further try to continue this procedure iteratively to higher
orders.

Let us  consider this procedure for the solution (\ref{KOansatz}).
In order to construct the quartic interactions we take four Hilbert spaces and
consider  the  Lagrangian
\begin{eqnarray} \label{LIBRSTQ}
{L} & = & \sum_{i=1}^4 \int d c_0^i \langle \Phi_i |\, Q_i \,|\Phi_i \rangle \ \\ \nonumber
  &&+ g( \int dc_0^1 dc_0^2  dc_0^3 \langle \Phi_1|
\langle \Phi_2|\langle \Phi_3||V_3 \rangle
+\int dc_0^1 dc_0^2  dc_0^4 \langle \Phi_1|
\langle \Phi_2|\langle \Phi_4||V_3 \rangle \\ \nonumber
&&+
\int dc_0^2 dc_0^3  dc_0^4 \langle \Phi_2|
\langle \Phi_3|\langle \Phi_4|V_3 \rangle
+
\int dc_0^1 dc_0^3  dc_0^4 \langle \Phi_1|
\langle \Phi_3|\langle \Phi_4||V_3 \rangle
+ h.c) \\ \nonumber
&& + g^2 (\int dc_0^1 dc_0^2  dc_0^3  dc_0^4  \langle \Phi_1| \langle \Phi_2|\langle \Phi_3| \langle \Phi_4|  |V_4 \rangle + h.c)
\end{eqnarray}
and the nonlinear gauge transformations
\begin{equation}
\delta | \Phi_i \rangle= (\delta_0 + \delta_1 + \delta_2) | \Phi_i \rangle
\end{equation}
where
\begin{equation}\label{BRSTIGT1Q0}
\delta_0 | \Phi_i \rangle  =  Q_i | \Lambda_i \rangle
\end{equation}
\begin{eqnarray} \label{BRSTIGT1Q1}
\delta_1 | \Phi_i \rangle &=&
- g(
\int
dc_0^{i+1} dc_0^{i+2}[(  \langle \Phi_{i+1}|\langle \Lambda_{i+2}|
+\langle \Phi_{i+2}|\langle \Lambda_{i+1}|) |V_3 \rangle] \\ \nonumber
&& \int
dc_0^{i+2} dc_0^{i+3}[(  \langle \Phi_{i+2}|\langle \Lambda_{i+3}|
+\langle \Phi_{i+3}|\langle \Lambda_{i+2}|) |V_3 \rangle] + \\ \nonumber
&&
\int
dc_0^{i+1} dc_0^{i+3}[(  \langle \Phi_{i+1}|\langle \Lambda_{i+3}|
+\langle \Phi_{i+3}|\langle \Lambda_{i+1}|) |V_3 \rangle]) \\ \nonumber
\end{eqnarray}
\begin{eqnarray}  \label{BRSTIGT1Q2-1} \nonumber
\delta_2 | \Phi_i \rangle &=&
{(-1)}^ig^2\int dc_0^{i+1}
dc_0^{i+2} dc_0^{i+3}[(  \langle \Phi_{i+1}|   \langle \Phi_{i+2}|\langle \Lambda_{i+3}| +
 \langle \Phi_{i+1}|   \langle \Phi_{i+3}|\langle \Lambda_{i+2}|+ \\
&& \langle \Phi_{i+2}|   \langle \Phi_{i+3}|\langle \Lambda_{i+1}|)
 |V_4 \rangle]
\end{eqnarray}

As it was for the case of cubic interactions, the Lagrangian (\ref {LIBRSTQ}) is invariant
up to zeroth order in the coupling constant i.e., under
 (\ref{BRSTIGT1Q0})
  since each separate BRST charge $Q_i$ is nilpotent. Further, the Lagrangian is
 invariant up to the fist order in the coupling constant i.e., under
(\ref{BRSTIGT1Q1}) provided
\be
(Q_i+Q_j+Q_k)|V_3 \rangle=0, \quad i \neq j  \neq k.
\ee

Let us now investigate the conditions on the quartic vertex. Up to
now our consideration has been completely general. In order to
simplify further our analysis, let us take a specific expression
(\ref{KOansatz}) for the cubic vertex where  cyclic symmetry
between indexes $i,j$ and $k$ is assumed. We also take diagonal
elements $Y_{ii}$ and $Z_{ii}$ equal to zero. From the cyclic
property of the cubic vertex it follows that the quartic vertex
obeys certain cyclic property as well.
 As  can be seen from (\ref{LIBRSTQ}) the quartic vertex $| V_4 \rangle$ should  change  sign under the cyclic permutation
of indexes $i,j,k,l$
\begin{equation}
| V_4(1,2,3,4) \rangle= - | V_4(2,3,4,1) \rangle=| V_4(3,4,1,2) \rangle= -| V_4(4,1,2,3) \rangle
\end{equation}
Our next step is to find a quartic vertex $| V_4 \rangle$ from the requirement of cancellation
of the terms of order $g^2$ in the variation of the Lagrangian. In particular,
a variation of  terms of the type
\be \label{TIP}
\tilde L=  g \int dc_0^1 dc_0^2  dc_0^3 \langle \Phi_1|
\langle \Phi_2|\langle \Phi_3||V_3 \rangle
\ee
under
\be
\tilde \delta_1 | \Phi_1 \rangle=
g \int dc_0^{2'} dc_0^{3'}[(  \langle \Phi_{2'}|\langle \Lambda_{3'}|
+\langle \Phi_{3'}|\langle \Lambda_{2'}|) |V_3 (1,2',3') \rangle]
\ee
should be compensated by the variation of the free part of the Lagrangian
under  (\ref{BRSTIGT1Q2-1}).
The latter gives
\be
(Q_1+Q_2+Q_3+Q_4) \langle \Phi_{i+1}| \langle \Phi_{i+2}|\langle \Phi_{i+3}| \langle \Lambda_{i+4}| |V_4 \rangle + 3 \quad permutations.
\ee
In order to evaluate the former consider the variation of a typical term
\bea
\delta \tilde L&=& g^2( \int dc_0^1 dc_0^2  dc_0^3
[ \langle V_3(1,2,3)| |\Phi_2 \rangle |\Phi_3 \rangle
\int dc_0^{2'} dc_0^{3'}
 \langle \Lambda_{2'}|
\langle \Phi_{3'}||V_3 (1,2',3') \rangle =  \nonumber \\ && +\int
dc_0^1 dc_0^2  dc_0^3 [ \langle V_3(1,2,3)| |\Phi_2 \rangle
|\Phi_3 \rangle \int dc_0^{2'} dc_0^{3'}
 \langle \Lambda_{3'}|
\langle \Phi_{2'}||V_3 (1,2',3') \rangle). \eea Let us integrate
first over the ghost variables which have an index $i=1$
evaluating explicitly the expression \bea &&\int dc_0^1 \langle
0_1 |c_0^1 exp (-Z_{ij} c^i b_0^j) \times exp( Z_{mn}c^{+m}b_0^n)
c_0^1 | 0_1 \rangle= \\ \nonumber &&=Z_{a'1} c^{+ a'} - Z_{a
1}c^a, \quad a'=2',3', \quad a=2,3. \eea For the integration over
the oscillator variable $\alpha_\mu^{+1}$ we use \be \langle 0_1 |
e^{Y_{1,i} \alpha^1_\mu p^i_\mu} e^{Y_{1,i} \alpha^{+1}_\mu
p^{i'}_\mu}| 0_1 \rangle = e^{Y_{1i'}Y_{1i} p^{i'}_\mu p^{i}_\mu}
\ee where we have used the conservation of momentum  $p^1_\mu = -
p^2_\mu - p^3_\mu= - p^{2'}_\mu - p^{3'}_\mu$. Collecting these
results
 one arrives to the equation for the quartic vertex\footnote{We shall keep implicit the cyclic symmetry of the quartic vertex  mentioned above.}
\begin{equation} \label{QEQ}
(Q_1+ Q_2 + Q_3 + Q_4) |V_4 \rangle =  18 (Z_{a'1} c^{+ a'} + Z_{a 1}c^{+a}) e^{Y_{1i'}Y_{1i} p^{i'}_\mu p^{i}_\mu} e^M c_0^2 c_0^3c_0^{2'} c_0^{3'} |0 \rangle_{232'3'}
\end{equation}
where
\begin{equation}\label{M}
M= Y_{a i} \alpha^{+ a}_\mu p_\mu^i + Y_{a' i'} \alpha^{+ a'}_\mu p_\mu^{i'} + Z_{ab}c^{+a}b_0^{b} +  Z_{a'b'}c^{+a'}b_0^{b'}
\end{equation}
with $i/i'=1,2,3/1,2',3'$, and  $a,b/a'b'=2,3/2'3'$ and the right
hand side of (\ref{QEQ}) can be written as a product of two cubic
vertices with appropriate gluing of two Hilbert spaces between
them \cite{Bengtsson:2006pw}. This is totally analogous to the
construction of the quartic vertex in OSFT by gluing two cubic
vertices with a star product $V_4 \sim V_3 \star V_3$. An obvious
ansatz for $|V_4 \rangle $ is
\begin{equation}\label{V4}
|V_4 \rangle = F(p) e^M c_0^2 c_0^3c_0^{2'} c_0^{3'} |0 \rangle_{232'3'}
\end{equation}
where the unknown function $F(p)$ is determined from (\ref{QEQ})
to be
\begin{equation}\label{F}
F(p)= \frac{18}{(p^2_\mu+ p^3_\mu)(p^{2'}_\mu+ p^{3'}_\mu )}.
\end{equation}
Obviously the full solution is given by acting with the above
given vertex on all non-cyclic permutations of the external states
\begin{equation}\label{LV4}
{\cal L}_4\sim \langle 1,2,3,4| V_4\rangle_s + \langle 1,3,2,4|
V_4\rangle_u + \langle 1,4,2,3| V_4\rangle_t \end{equation} where
each contribution has subscript indicating the massless pole on
the corresponding kinematic variable which comes from the
definition of $F(p)$ in (\ref{F}) and we use the standard
definitions for Mandelstam variables $s=(p_1+p_2)^2$,
$t=(p_1+p_4)^2$ and $u=(p_1+p_3)^2$. It turns out that the algebra
of gauge transformations remains abelian in this case and one can
check that the theory has vanishing four point functions. This is
related to the fact that the on-shell expression of
(\ref{KOansatz}) gives, for a HS state coupling to the other two
states, a current which is conserved identically and therefore
should not deform the gauge algebra \cite{Metsaev:2007rn}.
Recently in an updated version of \cite{Sagnotti:2010at} the
authors gave the completion of this vertex in the compensator
formalism, with the modification implied by the ansatz
(\ref{Bansatz1}) and (\ref{Bansatz2}). This is an extension of the
non-local vertex  (\ref{V4}) with the exponent $M$ having
additional terms due to the extra terms of ansatz
(\ref{Bansatz2}), compared to those of the ansatz for a cubic
vertex (\ref{KOansatz}) which we used to derive the quartic vertex
(\ref{V4}).

\setcounter{equation}0\
 \section{A short review of the BCFW method}\label{BCFWrev}

 All of the above attempts for constructing non-trivial
 interacting HS theories are plagued by several deficiencies. They
 deal with off-shell Lagrangians and therefore include non
 physical degrees of freedom which are eliminated only after proper gauge fixing.
   Also one can have various  physically equivalent
 descriptions related by field redefinitions. Moreover we have seen that
 although it is relatively straightforward to derive cubic Lagrangians which are    classically   non-trivial
 it is not so obvious how one can classify which of these Lagrangians can really lead to non-trivial theories beyond the cubic level.
 In a sense the construction of the algebra through the commutators of gauge transformations is
 the correct way to classify which interactions are non-trivial. Nevertheless, as we explained above field redefinitions and
 gauge trivial degrees of freedom make these methods rather cumbersome to implement even at the quartic level, see i.e. \cite{Bekaert:2005jf}.
 Extending these methods beyond the quartic level is a very challenging task.

 It is definitely desirable to have a method which deals
 directly with the physical degrees of freedom
 therefore allowing us to study the physical S-matrix of the
 theory. In any case all no--go theorems which forbid non-trivial
 HS interactions in 4-dimensions are formulated in the S-matrix
 language. Such a method has appeared recently \cite{Britto:2004ap, Britto:2005fq}
(BCFW)
 based upon the twistor formulation of gauge theories of Witten
 \cite{Witten:2003nn}. This method
 allows one to construct, under certain
 assumptions, higher point functions given the cubic ones.

The key point of BCFW \cite{Britto:2005fq} is that tree level
amplitudes constructed using Feynman rules are rational functions
of external momenta. Analytic continuation of these momenta on the
complex domain  turns the amplitudes into meromorphic functions
which can be constructed solely by their residues. Since the
residues of scattering amplitudes are, due to unitarity, products
of lower point on-shell amplitudes the final outcome is a set of
powerful recursive relations.

The simplest complex deformation involves only two external
particles whose momenta are shifted as
\begin{eqnarray}\label{BCFWshift1}
 \hp_i(z)= p_i - q z\ , \quad
 \hp_j(z)= p_j + q z \ .
\end{eqnarray}
Here $z \in \mathbb{C}$ and to keep the on-shell condition we need
$q \cdot p_{i} =q\cdot p_j= 0$ and  $q^2 = 0$. In Minkowski
space-time this is only possible for complex $q$. In particular,
if the dimensionality of space-time is  $d \geq 4$, we can choose
a reference frame where the two external momenta $p_i$ and $ p_j$
are back to back with equal energy scaled to 1
\cite{ArkaniHamed:2008yf} \be\label{mom} p_i  =
(1,-1,0,0,\ldots,0), \quad p_j = (1,1,0,0,\ldots,0), \quad q =
(0,0,1,i,0,\ldots,0)\ . \ee For gauge bosons the polarizations
under a suitable gauge can be chosen as \be\label{gpol}
\epsilon_i^+ = \epsilon_j^- = q, \quad \epsilon_i^- = \epsilon_j^+
= q^*, \quad \epsilon_T = (0,0,0,0,\ldots,0,1,0,\ldots,0) \ , \ee
where the $\pm$ superscripts correspond to the helicity of the
states which should not to be confused with the usual light-cone
notation for massless states.  These vectors and the  $D - 4$
different polarizations $\epsilon_T$ form a basis in the
transverse directions \cite{ArkaniHamed:2008yf}.

Under the deformation (\ref{BCFWshift1}) the polarizations should
become
 \be\label{gpolshift} \he_i^+(z) =
\he_j^-(z) = q, \quad \he_i^-(z) = q^* - z p_j, \quad \he_j^+(z) =
q^* + z p_i,
\ee
$$
\he_T(z) = (0,0,0,0,\ldots,0,1,0,\ldots.,0)
$$
so that they remain orthogonal to the shifted momenta. In {\em
four dimensions} we can use spinor representation of momenta and
polarizations (for spin 1 states)
\begin{eqnarray}\label{spinor}
&&p^\mu = \lambda^a (\sigma^\mu)_{a\dot a}\tilde\lambda^{\dot
a}\nonumber \\
 &&\epsilon^+_{a\dot a} = \frac{\mu_a\tilde\lambda_{\dot
a}}{\langle \mu ,\lambda \rangle},  \qquad  \epsilon^-_{a\dot a}
=\frac{\lambda_a\tilde\mu_{\dot a}}{[\tilde\lambda , \tilde\mu]} \\
&& \langle \mu ,\lambda \rangle \equiv \mu_a \lambda_b
\epsilon^{ab} \qquad [\tilde\lambda , \tilde\mu] \equiv
\tilde\mu_{\dot a} \tilde\lambda_{\dot b} \epsilon^{\dot a \dot
b}\nonumber
\end{eqnarray}
with $\mu_a$ and $\tilde\mu_{\dot a}$ arbitrary reference spinors.
Polarizations of Higher Spin states  are given by products of the
polarizations for spin 1
\begin{equation}\label{polten}
\epsilon^+_{a_1 \dot{a}_1 \dots a_s \dot{a}_s} = \prod_{i=1}^s
\epsilon ^+_{a_i \dot{a}_i} \qquad \epsilon^-_{a_1 \dot{a}_1 \dots
a_s \dot{a}_s} = \prod_{i=1}^s \epsilon ^-_{a_i \dot{a}_i}
\end{equation}  The expressions above for the BCFW
shifted momenta and polarizations correspond to the following
shift on the spinors
\begin{equation}\label{shift}
\hat{\lambda}_a^{(i)}(z) = \lambda_a^{(i)} + z\lambda_a^{(j)}, \qquad
\hat{\tilde\lambda}_{\dot{a}}^{(j)}(z) = \tilde\lambda_{\dot{a}}^{(j)} -
z\tilde\lambda_{\dot{a}}^{(i)}
\end{equation}

A general amplitude, which after the deformation becomes a
meromorphic function
 ${\cal M}_n(z)$,
 will have simple poles for those values of $z$
where the propagators of intermediate states go on shell (on the
complex plane) \be\label{BCFWpole} \frac{1}{P_J(z)^2} =
\frac{1}{P_J(0)^2 - 2 z q \cdot P_J}\ .
 \ee
 We denote by $P_J= \sum_{i \in J} p_i$ the total momentum of the
 intermediate state which connects the two sub-amplitudes
 in which ${\cal M}_n(z)$ factorizes.
 The undeformed amplitude
can be computed using Cauchy's theorem:
\begin{equation}\label{BCFWrel1}
{\cal M}_n (0) = {1\over 2\pi i}\oint_{z=0} \frac{{\cal M}_n(z)}{z} dz = -
\left\{\sum \mathrm{Res}_{z=\textrm{finite}} + \mathrm{Res}_{z=
\infty} \right\} \ .
\end{equation}

As already stated, the residues at finite locations on the complex
plane are necessarily, due to unitarity, products of lower point
tree level amplitudes which are however computed at complex on-shell momenta.
The residue at infinity can have a  similar interpretation is some
special cases \cite{Feng:2009ei,Feng:2010ku} but in general it
cannot be written as product of lower point amplitudes. In most
cases involving gauge bosons and/or gravitons there is an
appropriate choice of the deformed external polarizations such
that under a shift of the type (\ref{BCFWshift1}), ${\cal M}_n(z)$
vanishes in the limit $z \rightarrow \infty$. In these cases the
BCFW relation takes the simple form
\begin{equation}\label{BCFWrel2}
{\cal M}_n(1, \ldots, n) = \sum_{r,h(r)}
\sum_{k=2}^{n-2}
\frac{{\cal M}_{k+1}(1,2,\ldots, \hat{i}, \ldots, k, \hat{P}_r)
{\cal M}_{n-k+1}(\hat{P}_r, k+1, \ldots, \hat{j}, \ldots,
n)}{\left(p_1 + p_2 + \ldots + p_k \right)^2 } \ ,
\end{equation}
where the hat is used for the variables which are computed at the
residue of the corresponding pole (\ref{BCFWpole}) and satisfy the
physical state condition. Since we have made a complex deformation
the hatted variables correspond to complex momenta, unlike the
momenta of the rest of the external particles. The sum in $r$ is
over all different particles of the theory which can propagate in
the intermediate channel and the sum in $h(r)$ is over their
helicities.

In \cite{Benincasa:2007xk} a very interesting criterion was
derived in order to classify, in four dimensions, which theories
are constructible under BCFW deformations. The criterion has been
stated explicitly for the four-point function and it is a
necessary condition for a theory to have zero residue at infinity,
i.e. to be constructible.  Say we denote by ${\cal
M}^{(i,j)}(z)$ the four-point function under deformation of
particles $i$ and $j$. Assume further on that the helicities $h_2$
and $h_4$ are negative while $h_1$ is positive.  The criterion
advocates that
\begin{equation}\label{BC1}
{\cal M}_4^{(1,2)}(0)={\cal M}_4^{(1,4)}(0)
\end{equation}
This is highly non-trivial since the usual Feynman analysis
construction of the four-point function requires adding diagrams
from three possible channels while each BCFW deformation can use
only two channels of exchanged particles. For $M^{(1,2)}(0)$
diagrams where particles 1 and 2 go to an intermediate state do
not lead into poles on the complex plane i.e. $1/
(p_1(z)+p_2(z))^2=1/ (p_1+p_2)^2$. So in $M^{(1,2)}(z)$ only poles
from the $t,\ u$ channels on the complex z-plane will contribute.
Viceversa for $M^{(1,4)}(0)$ only poles from the $s, \ u $
channels on the complex-z plane will contribute. So the {\it
crossing symmetry} condition (\ref{BC1}) is a highly nontrivial
constrain.

 After this
short review of the BCFW method we will proceed by considering  some
examples of theories of massless HS fields which are based on
recent studies of the high energy limit of string theory
\cite{Fotopoulos:2007nm, Taronna:2010qq}. As was shown in
\cite{Boels:2010bv, Cheung:2010vn}, tree level string amplitudes
for appropriate regimes of the Mandelstam variables
 vanish when $z \to \infty$. Therefore one expects that these amplitudes
 also satisfy
 recursive relations similar to the field theoretic ones (\ref{BCFWrel2}).The key point, as we will
explain below, is that the methods employed in \cite{Boels:2010bv,
Cheung:2010vn} using the Pomeron technique of \cite{Brower:2006ea}
suggest that the naive tensionless limit $\apr \to \infty$ does
not spoil the good behavior of string amplitudes for large complex
deformation of the momenta. Therefore one would expect that if a
consistent high energy limit of string theory indeed exists then
the theory would have to be BCFW constructible in the sense above.
A  study for some couplings was given in \cite{Benincasa:2007xk}
but it was limited  to self interactions and did not attempt to
address  either a complete theory or  an example where all cubic
couplings are given. We will try to shed some light on recently
proposed theories of interacting HS and derive useful conclusions.

\setcounter{equation}0\
\section{BCFW for Higher Spin theories}\label{BCFWHS}
For any Lorentz invariant theory of massless particles in
4-dimensions it is straightforward to see that the on-shell cubic
amplitudes vanish. Extending though some of the momenta in the
complex plane we can easily derive that the most generic cubic
amplitude is given by \cite{Benincasa:2007xk}
\begin{equation}\label{cubicamp}
{\cal M}_3(\{\lambda^{(i)},\tilde\lambda^{(i)},h_i\}) = \kappa_H
\langle 1,2\rangle^{d_{3}}\langle 2,3\rangle^{d_{1}}\langle
3,1\rangle^{d_{2}}+ \kappa_A [1,2]^{-d_{3}}[ 2,3]^{-d_{1}}[
3,1]^{-d_{2}},
\end{equation}
where we have defined $d_1=h_1-h_2-h_3, \ d_2=h_2-h_1-h_3, \
d_3=h_3-h_1-h_2$ and the coupling constant $k_H$ ($k_A$) is
required to give the right dimension to the part of the amplitude
which is holomorphic (antiholomorphic) in the spinor variables
$\lambda^{(i)}$ ($\tilde{\lambda}^{(i)}$). Imposing that ${\cal
M}_3$ has the correct physical behavior in the limit of real
momenta.  shows that if $d_1+d_2+d_3$, which is equal to
$-h_1-h_2-h_3$, is negative (positive) then we must set
$\kappa_H=0$ ($\kappa_A =0$) in order to avoid a singularity when
any two (anti)holomorphic spinors become linearly dependent. The
case when $h_1+h_2+h_3 = 0$ is more subtle since both pieces are
allowed.

Under the conditions above one can show that ${\cal
M}_4^{(1,2)}(0)$ is given by the expression
\cite{Benincasa:2007xk}
\begin{eqnarray}\label{M12}
{\cal M}_4^{(1,2)}(0) = && \sum_{h > {\rm
max}(-(h_1+h_4),(h_2+h_3))} \big(
\kappa^A_{1-h_1-h_4-h}\kappa^H_{1+h_2+h_3-h}
\frac{(-P_{3,4}^2)^h}{P_{1,4}^2}\left(\frac{[1,4][3,4]}{[1,3]}\right)^{h_4}
 \nonumber \\
&& \left(\frac{[1,3][1,4]}{[3,4]}\right)^{h_1}\left(\frac{\langle
3,4\rangle}{\langle 2,3\rangle\langle
2,4\rangle}\right)^{h_2}\left(\frac{\langle 2,4\rangle}{\langle
2,3\rangle\langle 3,4\rangle}\right)^{h_3}\big)  \\ \nonumber
&&+ \sum_{h
> {\rm
max}(-(h_1+h_3),(h_2+h_4))}\!\!\!\!\!\!\!\!(4\leftrightarrow
3).\nonumber
\end{eqnarray}
where $P_{i,j}= p_i +p_j$. This is the final formula we will need
to impose the constructibility criterion for the massless HS
vertices we wish to examine. The subscript of the coupling
constants denotes their mass dimension.

We will consider a theory with cubic couplings of two scalars and
one HS field to study the $2 \to 2$ amplitude for scalars. Notice
that usually in the BCFW method we consider external states which
are gauge bosons. The reason is that the polarization tensors in
(\ref{spinor}), with polarization assignments as stated above,
lead to factors  $1/z$ and are important for the vanishing residue
of ${\cal M}(z)$ at infinity. In our case though it is obvious
that since the dimension of the couplings decreases with the spin,
the behavior of individual Feynman diagrams becomes more and more
divergent and therefore only subtle cancellations among Feynman
diagrams can give a well behaved amplitude at infinite complex
momenta. So we can work with the elementary process of scalar
scattering which encodes all physics relevant to the theory.

We can write the cubic interaction for two scalars and one
arbitrary HS triplet \cite{Fotopoulos:2009iw}.  In order to
compute the S-matrix elements for triplets one would need the
propagator of the irreducible states they are composed of. This
will introduce the normalization factors of
\cite{Fotopoulos:2009iw}. In order to avoid technical
complications which involve such factors and do not alter the
physical properties we want to study, we will simplify our
interacting Lagrangian and we will discuss a Lagrangian coupling
of irreducible HS fields $\Psi_h$ to two  scalars \cite{Berends:1985xx}
\begin{equation}\label{ILW}
{\cal L}_{int}^{00s}=   \kappa^{1-h}  N_{h} \ {\Psi_h^{\mu_1\dots \mu_h}
J^{1;2}_{h; \mu_1\dots \mu_h}\over h!} + \ h.c.\,,
\end{equation}
where $\kappa$ is a dimensionful constant needed in order to give
the right dimension to the Lagrangian and we have assumed that the
free action is normalized to give the properly normalized
propagator for the fields with residue one. The interaction is
proportional to the coefficients $Y_{ij}$ of (\ref{KOsolution})
but we shall not give the exact relation between them since our
discussion will be more general than the specific solution.
Moreover $N_{h}$ is an undetermined constant which probably gets
fixed once we have the fully consistent, interacting HS theory to
all orders in the coupling constant.
 The currents are defined as
\begin{equation}\label{Jp}
J^{1;2}_{h;\mu_1\dots \mu_h}=\sum_{r=0}^{h} \ \bn{h}{r} \ (-1)^r \
(\partial^{\mu_1} \dots
\partial^{\mu_r}
\phi_1)\ (\partial^{\mu_{r+1}} \dots
\partial^{\mu_{h}}\phi_2)\,
\end{equation}
For example in the theory of \cite{Fotopoulos:2007nm} from the
high energy limit of bosonic SFT one has $N_{h}=\sqrt{h!}$ for the
top spin component of each triplet. Of course using the
decomposition method in \cite{Fotopoulos:2009iw} the lower spin
components of each triplet will have different normalizations
dependent on the spin of the triplet they descend from. In
principle in order to discuss on safe ground the high energy limit
of SFT one should include all subleading Regge trajectories, mixed
symmetry fields, while decomposing them into irreducible modes.
For the purpose of our discussion we will concentrate into two
main examples of couplings, which have been studied in
\cite{Bekaert:2009ud, Sagnotti:2010at}, with a simple spin
dependence. The first example we will call {\it string theory}
coupling has $N_{h}=\sqrt{h!}$ and it is  the same as in
\cite{Taronna:2010qq} derived from string theory. The second one
leads to spin independent coupling constants and we will call it
{\it field theory} coupling. The normalization constant has a
stronger growth with spin and is given by $N_{h}=h!$. This is
unlike the softer {\it string theory} normalizations which lead
into coupling constants which drop with increasing spin.

From the Lagrangian terms in (\ref{ILW}) we can infer the three
point on-shell amplitudes we will need for our test of the BCFW
construction. We remind the reader that these three-point
amplitudes make sense only for complex external momenta, since
otherwise 4-dimensional Lorentz invariance requires that all
on-shell three point amplitudes for massless states vanish. We can
use directly (\ref{cubicamp}) to infer that for particles 1 and 2 being
scalars and the particle 3 being a spin h particle
\begin{eqnarray}\label{M00h}
M^{00+h}_3({\lambda_i, \tilde{\lambda}_i, h_i})&=&
k^A_{1-h}(h) {[2,3]^{h} [3,1]^{h} \over [1,2]^{h}} \nonumber \\
M^{00-h}_3({\lambda_i, \tilde{\lambda}_i, h_i})&=& k^H_{1-h}(h){
\langle 2,3 \rangle ^h \langle 3,1 \rangle ^h \over \langle 1,2
\rangle ^{h}}
\end{eqnarray}
where the couplings of the two amplitudes are given by
\begin{equation}\label{k}
k^A_{1-h}(h)=k^H_{1-h}(h)= \kappa^{1-h} {N_h \over h!}.
\end{equation}
Notice that the amplitudes must be symmetric under exchange of the
two scalars. If the theory has only one scalar field or if we
assume an abelian theory the spin h is restricted only to even
values. We can assume couplings which have Chan--Paton factors for
the scalars in such a way that we can include odd-spin couplings
to our computations. In a way we will be computing colour ordered
amplitudes. The total scattering amplitude will be given by the
sum of the colour ordered terms with their corresponding color
factors. Obviously if we go to the abelian case, i.e., when colour
factors are trivial, only couplings to even HS fields will
survive. As an example for the cubic amplitude there are two
different color orderings given by exchange of particles 1 and 2
in (\ref{M00h}). If we have a non-abelian theory they are
nonequivalent but in the abelian case they will simply add up and
since they have opposite signs for couplings to odd spin HS fields
they will cancel. In what follows in order to avoid incorporating
Chan--Paton factors in our notation, since it is an unnecessary
complication, we will focus our discussion on the abelian case.
Nevertheless since at several points we would like to compare our
results with the explicit Feynman diagrams of
\cite{Bekaert:2009ud}, we will assume that there are at least two
scalar fields or a complex one. In other words we will consider
scalar fields which carry some charge. This will allow couplings
of odd HS fields.

In order to derive the exact form of the couplings one would need
to substitute (\ref{spinor}, \ref{polten}) in (\ref{ILW}) where it
turns out, as expected due to gauge invariance, that the
dependence on the reference spinors $\mu_i$ drops out. Actually in
this case we can easily write down a coupling for $h=0$. Remember
that this is the case where $h_1+h_2+h_3=0$ and both a holomorphic
and antiholomorphic piece are allowed. In this case the situation
is rather trivial and we can take unambiguously $h\to 0$ in
(\ref{M00h}) and we end up with a constant $M^{000}=  N_0 \kappa$.

 Before we proceed with the
study of BCFW recursion relations for these two kinds of couplings
we would like to make a few comments for the string theory
coupling. In string theory we expect the four point function to
take the familiar Veneziano type amplitude behavior. This means
that four point string amplitudes have Gamma function prefactors.
A BCFW shift of the momenta results into shifts of two of the
three Mandelstam kinematic invariants and naive Stirling
approximation would give badly behaved amplitude for some region
of complex infinity. Nevertheless the fact that these Gamma
functions appear into Beta function combinations allows one to see
that indeed the four point function is a well behaved polynomial
of $1/z$ at infinity \cite{Boels:2008fc,Boels:2010bv}. For higher
point functions it turns out it is more straightforward to use the
Pomeron technique of \cite{Brower:2006ea} to study the behavior of
amplitudes under BCFW deformations \cite{Cheung:2010vn}. The
typical behavior is
\begin{equation}\label{BCFWbeh}
{\cal M}(z) \sim z^{n+1+ \apr P_{12}^2}
\end{equation}
where $n$ is the level of the external string states which are being
deformed and $P_{12}= p_1 +p_2$. We see that choosing $n+1+\apr
P_{12}^2<0$ we can conclude that all open bosonic string
amplitudes are constructible. This might require an analytic
continuation of the external states momenta outside the physical
region but it is  standard practice with string amplitude
computations. This analytic continuation should not be confused
with the complex momentum deformation. Moreover, the formulas we
gave above in (\ref{M12}) contain no assumptions about reality of
the undeformed spinors.  Now it is obvious that taking $\apr \to
\infty$ naively makes things only better \footnote{ A possible
caveat to this argument is the following. The general form in
(\ref{BCFWbeh}) is
\begin{equation} {\cal M}(z) \sim z^{n+1+ \apr P_{12}^2} \ {\cal
H}(\apr, p_i, z)
\end{equation} We have two limits to take, $z\to \infty$ and $\apr \to \infty$.
The function ${\cal H}(\apr, p_i, z)$ is a polynomial in inverse
powers of $z$, therefore the limit $z\to \infty$ should pose no
problem. Since the leading term dependent to $z$ , $z^{n+1+ \apr
P_{12}^2} $, can only behave better with $\apr \to \infty$ for
suitable $P_{12}^2$ as explained above, we can assume taking first
$z\to \infty$ and then the high energy limit. Even if ${\cal
H}(\apr, p_i,z)$ is divergent at the high energy limit this would
not be a problem if we assume that the two limits commute. To put
it another way the BCFW shift corresponds to taking a large
hierarchy between kinematic variables of a given scattering
amplitude i.e. $s\gg t$. Taking the high energy limit of the
theory, which means $\apr s, \apr t \gg 1$, but keeping the
original hierarchy between the kinematic variables should not be a
problem and our arguments, about BCFW constructibility of the
theory in the high energy limit, should hold.}. But we know that
the naive limit might lead to an non-interacting theory as well
\cite{Gross:1987ar} so it might be a bit too naive to conclude
anything more than good behavior of the amplitudes at infinity. It
makes more sense to assume that if an interacting theory indeed
exists in the high energy limit of string theory, then the string
length reaches a critical value, $\apr \to \apr_{cr}$, rather than
infinity. The Pomeron argument seems still valid so if indeed a
consistent high energy limit of string theory does exist, we would
expect that it passes the BCFW test (\ref{BC1}). Finally a careful
examination of the saddle point method, used to derive the Pomeron
operators which led to the aforementioned behavior, does not show
any pathology under the tensionless limit.

One can immediately see that there might be potential problems
trying to apply the BCFW procedure to interacting massless HS
theories like those discussed in the previous sections. The
exponential type vertex in (\ref{V4}), unlike the Veneziano
amplitude, does not have a good behavior through all complex
infinity under BCFW deformations and we will have to discuss novel
features like inclusion of Pomeron-like states in the theory in
order address this problem.

In the next
two subsections we will study the BCFW test of
\cite{Benincasa:2007xk} for two potentially interesting cases:
{\it string and field theory couplings}.

\subsection{BCFW for field theory couplings}\label{BCFWft}
In this case we need to use the coupling constants $N_h=h!$. We
will consider the scattering process $\phi(p_1) \ \phi(p_2) \to
\phi(-p_3) \ \phi(-p_4)$ for two  scalars with the same charge as in
\cite{Bekaert:2009ud}. To this end we need to apply equation (\ref{M12}) using
(\ref{M00h}). A straightforward computation gives
\begin{equation}\label{M12FT}
{\cal M}_4^{(1,2)}(0)= \sum_{h \in  \ \mathbb{Z}} \kappa^{2-2h}
(-P^2_{3,4})^h \left( {1\over P^2_{1,4}} + {1\over P^2_{1,3}}
\right).
\end{equation}
Using the standard definitions for Mandelstam variables
$s=(p_1+p_2)^2$, $t=(p_1+p_4)^2$ and $u=(p_1+p_3)^2$ in $(+,-,-,-)$ signature we can write
the result as
\begin{equation}\label{M12FTb}
{\cal M}_4^{(1,2)}(0)= \kappa^2 {1 \over 1+ \kappa^{-2} s} \left(
{-s\over t u}\right) .
\end{equation}
On the other hand we can exchange labels 2 and 4 in the expression
(\ref{M12FT}). In this case only the u-channel (with an
intermediate propagator $1/P^2_{1,3}$) contributes due to charge
conservation. The s-channel amplitude, (with an intermediate
propagator $1/P^2_{1,2}$,) is not allowed in this process but
should be used in the process $\phi^*(p_1) \ \phi(p_2) \to
\phi(-p_3) \ \phi^*(-p_4)$, where $\phi^*$ is the antiparticle of
$\phi$. Finally we get
\begin{equation}\label{M14FT}
{\cal M}_4^{(1,4)}(0)= \kappa^2 {1 \over 1+ \kappa^{-2} t} \left(
{1\over  u}\right).
\end{equation}
We see immediately that the two results do not match. This tells
us that either the theory is not constructible or there is some
pathology in its definition. Computing the explicit Feynman
diagrams as in \cite{Bekaert:2009ud} for {\it field theory
couplings} of the form in (\ref{ILW}) shows that indeed we have a
function with a pole at finite radius on the complex kinematic
plane and that the position of the pole depends on the kinematics.
The t-- channel of the four point amplitude has the form
\begin{equation}\label{CE}
{\cal M}^t_4 \sim {\kappa^{2}\over t} \left( {1 \over 1 +
{\kappa^{-2}\over 4} (\sqrt{s} +\sqrt{-u})^2}+ {1 \over 1 +
{\kappa^{-2} \over 4}(\sqrt{s} -\sqrt{-u})^2}-1\right).
\end{equation}
The full amplitude is given by summing the u-channel contributions
as well. The full expression should be compared to the BCFW
result. Looking at the expression above for the t-channel suggests
two things. The first one is that the original series of massless
exchanges, either in the Feynman method or the BCFW one, has a
finite radius of convergence. Outside this radius the series is
defined through analytic continuation just as $\sum_s z^s= {1\over
1-z}$. Thus it is fairly obvious to see that a BCFW shift as in
(\ref{shift}) results in a vanishing amplitude at infinite complex
deformation which means that the theory should be constructible.
The second point is that we have a pole which depends on the
kinematic variables. This suggests that there should be some
extended object in the theory since the amplitude blows up for
specific impact parameters and angles \cite{FotopoulosUn,
Taronna:2010qq}. The BCFW computation tells us the same thing
through its failure of crossing symmetry under BCFW deformations
which suggests that at some finite distance on the complex
kinematic variables plane the massless theory has some ingredient
missing in its definition \footnote{ The final result in
(\ref{M12FTb}) has a massive pole outside the physical region i.e.
for $s<0$. In the physical region though it is a form factor for
an apparent massless exchange of particles which depends on the
center of mass energy for (\ref{M12FTb}) and angle of scattering
for (\ref{M14FT})}. Actually from the explicit expression in
(\ref{CE}) we can be a bit more precise. The amplitude is a
meromorphic function of the kinematic variables. All dependence on
the square roots of the kinematic variables disappears once we add
the two terms together. Therefore this function should be
determined through its poles on the complex plane. Under a BCFW
shift (\ref{shift}) $\hat{\lambda}_a^{(1)}(z) = \lambda_a^{(1)} +
z\lambda_a^{(2)}, \ \hat{\tilde\lambda}_{\dot{a}}^{(2)}(z) =
\tilde\lambda_{\dot{a}}^{(2)} - z\tilde\lambda_{\dot{a}}^{(1)}$
the Mandelstam variables shift as
\begin{equation}\label{Mshift}
\hat{s}= s \ , \quad \hat{t}= t-z \langle 2, 3\rangle [1, 3]\ ,
\quad \hat{u}= u+z \langle 2, 3\rangle [1, 3].
\end{equation}
Then  one can easily verify that there are three poles on the
complex z-plane for the t-channel contribution and three for the
u-channel respectively. The one which comes from the massless
t-pole of (\ref{CE}) is the one whose $Res ({\cal M}(z)/z)$
reproduces the t-channel pole of (\ref{M12FT}). On this pole the
amplitude factorizes into two scalar three point amplitudes which
are however dressed with some form factors. The other two poles of
(\ref{CE}) should come from the "extended object" poles which
contribute the form factors of (\ref{M12FTb}). To get the
u-channel pole of (\ref{M12FT}) we will need to compute the
residues of the u-channel amplitude ${\cal M}_4^u$ in a similar
manner. We can repeat the exercise for scattering of real scalars.
This will require, as we mentioned before, that in (\ref{M12FT})
we will sum over even spins only and the same for the Feynman
diagrams computation in \cite{Bekaert:2009ud}.

\subsection{BCFW for string theory couplings}\label{BCFWst}
In this section we will try to repeat the analysis above for the
{\it string theory} couplings. One should keep in mind that we are
not certain about the exact nature of the four-point function at
the high energy limit of string theory. The standard result of
\cite{Gross:1987ar} gives a behavior which does not seem that it
can be reproduced by a local field theory \footnote{ At least
perturbatively non-local in the sense of \cite{Bekaert:2010hw}}.
As pointed out in \cite{Taronna:2010qq}, for a Feynman diagram
computations of the full 4-point function the subleading Regge
trajectories of their vertex would be necessary. But this by
itself would not be enough since a non--local quartic vertex would
be required in order to allow one to reproduce the local and
unitary result of \cite{Gross:1987ar}. Moreover, the high energy
behavior of \cite{Gross:1987ar} is proportional to the fixed angle
scattering of string amplitudes
\begin{equation}\label{Gross}
{\cal M}_4 \sim e^{-\apr s \log{\apr s} -\apr t \log{\apr t}-\apr
u \log{\apr u}}\ .\end{equation} The result above is definitely
divergent under any BCFW deformation. But on the other hand as we
pointed out the Pomeron arguments suggest that under $\apr \to
\infty$ high energy string amplitudes remain constructible.
Instead of relying on the standard result of \cite{Gross:1987ar}
we will take a more pedestrian approach and assume as mentioned
before that the theory reaches a critical string length were the
theory is interacting. We will then try to construct amplitudes
via BCFW and apply the criterion (\ref{BC1}). This will fail as we
shall shortly see and we will suggest a possible explanation for
the problem.

Lets first discuss the amplitude computed using current exchanges
for massless HS states in \cite{Bekaert:2009ud} for the
$t-channel$ for the process of two same charge scalar scattering
as in subsection \ref{BCFWft}
\begin{equation}\label{CEst}
 {\cal M}_4^t =
-\frac{\kappa^{2}}{t}\,\left[\,
2\,\exp\!\Big(-\frac{\kappa^{-2}}{4}\,(s-u)\Big)\,
\cosh\!\Big(\,\frac{\kappa^{-2}}{2}\,\sqrt{-su}\,\Big)-1\,\right]
.
\end{equation}
This amplitude has a structure very similar with the one of the
vertex (\ref{V4}). In the Regge regime with $s\gg t$, which
implies in the massless case $s\sim -u$, we easily derive
\begin{equation}\label{CEstregge}
{\cal M}_4^t \sim -\frac{\kappa^{2}}{t}\
\exp\!\left(-\kappa^{-2}\,s\right) \ . \end{equation} For
$s\sim-u\sim q z$ this corresponds to the BCFW shift
(\ref{Mshift}) of momenta $p_1, \ p_4$ in (\ref{CEst}) and
certainly does not vanish for all $z$ in complex infinity. But the
statement for BCFW constructibility requires that the full
amplitude vanishes at complex infinity and not individual Feynman
diagrams. The u-channel contribution cannot cancel that of the
t-channel at complex infinity as one can easily check. It is
plausible though that the BCFW deformation of the corresponding
quartic vertex (\ref{V4}) gives a contribution which cancels the
divergence from the exchange diagram (\ref{CEstregge}).  The
actual form of the quartic vertex contribution to the four point
function is of the form
\begin{equation}\label{M4cont}
{\cal M}_4^{t;contact} \sim {18 \over Y t} \
\exp\!\Big(-Y\,(s-u)\Big)
\end{equation}
where we have assumed that all coupling constants of the theory
are determined in terms of one scale $Y_{ij}\sim Y(\kappa)$. One
can compute explicitly the quartic vertex contribution with the
exact normalization and of course determine the exact form of
$Y(\kappa)$. If we determine the exact form of $Y$ and
normalization of the contribution in (\ref{M4cont}) we might
cancel, under BCFW shift, the problematic behavior in
(\ref{CEstregge}). This would of course imply that the theory is
constructible. Nevertheless at this point we will not try to
verify if such a cancellation is true since, as we shall see
bellow, it does not seem possible to apply the criterion
(\ref{BC1}) successfully. This conflict between the apparent
constructibility of the theory and failure of (\ref{BC1}) seems to
imply that we need to make further assumptions for the structure
of the theory.

Before we proceed further it would be useful to discuss the role
of mixed symmetry fields to the arguments above. We pointed out in
section \ref{QUAR}, that if one starts with the cubic vertex for
the HS fields given in (\ref{KOansatz}) and then adds the
contribution of the quartic vertex (\ref{V4}) the full four point
function vanishes. Presumably, performing similar computation the
four point function for the theory where the cubic vertex is given
by (\ref{Bansatz1}), could lead to a non-trivial four-point
function.
We would like to point out  that
the couplings of reducible mixed
symmetry HS fields to two scalars are identical to those derived
the vertex of (\ref{KOansatz}) since the currents built from the
two scalars are totally symmetric. Decomposing the reducible mixed symmetry
fields \footnote{ Irreducible mixed symmetry fields give zero coupling to
a totally symmetric current.} to their symmetric parts results into various
HS symmetric
fields with multiplicities dependent on the Young tableaux of the
mixed symmetry field and moreover on the generalized triplet they
descend from.
 So in order to make precise statements about the
behavior of scattering amplitudes at infinite complex momenta, one
would need to compute these couplings for all generalized triplets
and show that the full amplitude constructed using Feynman
diagrams vanishes for infinite complex momenta. Alternatively one
could use the method of \cite{ArkaniHamed:2008yf}. But unlike spin
one and spin two gauge theories, where the Lagrangian analysis
leads directly to a well-behaved theory at complex infinity, here
it would be subtle cancellations between interactions of all HS
modes which can grant a soft behavior. So we would need at least
some knowledge of the full HS Lagrangian for all mixed symmetry
fields.

In any case if such an analysis would be possible and the theory
turns out to be well-behaved at complex infinite momenta, then one
would expect that the theory with such cubic vertex would pass the
test (\ref{BC1}) irrespective of the precise form of the quartic
vertex. This is actually the power of BCFW consistency conditions,
that if the test fails, either there is a missing ingredient of
the theory or the BCFW deformed amplitude does not fall fast
enough and there is a boundary contribution to the BCFW relations and
the method is not applicable. We used the Pomeron analysis to
argue that the boundary contribution is vanishing for the
tensionless limit of string amplitudes. So we will use as working
hypothesis that this holds true and the BCFW recursion relations
should apply.

Let us look at the BCFW construction of the four point function
for scalar fields for $N_h=\sqrt{h!}$.
\begin{equation}\label{M12ST}
{\cal M}_4^{(1,2)}(0)= \sum_{h \in  \mathbb{Z}}
{\kappa^{2-2h}\over h!} (-P^2_{3,4})^h \left( {1\over P^2_{1,4}} +
{1\over P^2_{1,3}} \right)
\end{equation}
\begin{equation}\label{M12STb}
{\cal M}_4^{(1,2)}(0)= \kappa^2 e^{ -\kappa^{-2} s} \left(
{-s\over t u}\right)
\end{equation}
On the other hand we can exchange label 2 and 4 and we get
\begin{equation}\label{M14ST}
{\cal M}_4^{(1,4)}(0)= \kappa^2 e^{ -\kappa^{-2} t} \left( {1\over
 u}\right)
\end{equation}
where as mentioned in subsection \ref{BCFWft} only the u-channel
contributes for this scattering process. We immediately see a
couple of problems. The first problem is that neither of the two
expressions of the amplitude does allow for soft behavior under
all possible BCFW shifts.
  The second point is that the results in (\ref{M12STb}) and (\ref{M14ST}) would be rendered
consistent if the exponential function $e^{-\kappa^{-2}s}$ could
become i.e. $\left( {1\over \kappa^{-2}s}\right)$, and similar for
$e^{-\kappa^{-2}t}$, therefore fixing the problematic exponential
behavior and allowing us to pass the test (\ref{BC1}). It is
obvious that such a thing is not possible since the expression in
(\ref{M12ST}) is a Laurent series around $s=0$ and cannot give a
function with a pole at $s=0$. Could we get a result like in the
previous subsection with a pole at a some other location on the
complex $s-$plane? As was pointed out in \cite{Taronna:2010qq} and
we commented above, only the totally symmetric part of mixed
symmetry fields couples to the currents (\ref{Jp}) and therefore
leads to the same couplings as the totally symmetric fields. One
expects of course that the normalization constants $N_h$ of these
terms will become dependent on the Regge trajectory each couplings
originates from. So it is plausible that $N_h$ might grow stronger
with spin than $N_h\sim \sqrt{h!}$ of the totally symmetric fields
leading to a behavior similar to (\ref{M12FTb}) . This point
deserves further investigation but the obstacle is still the
knowledge of the full fledged HS theory. In any case, even if this
is possible then the discussion of the previous subsection would
apply and it would mean that one might have to include extended
objects in the theory. Therefore the theory is not complete with
the massless HS fields on their own. The behavior suggested in
\cite{Gross:1987ar} seems beyond reach since we cannot expand
(\ref{Gross}) in terms of massless exchanges.

In order to get an idea of what might be the problem we should
remind ourselves of some important features of the prototype case,
the Veneziano amplitude. It is the beta function expansion of the
Veneziano amplitude,
\begin{equation}\label{beta}
{\cal M}_4 (s,t)= -\sum_{n=0}^{\infty}
{(\alpha(s)+1)(\alpha(s)+2)\dots(\alpha(s)+n)\over n!}\  {1\over
\alpha(t)-n}
\end{equation}where $\alpha(s)= \apr s + \alpha_0$ the Regge
trajectory, which gives it the nice properties of dual models. The
massive poles are responsible for the dual nature of string
amplitudes. The result above in (\ref{beta}) can be equivalently
expanded in s--channel poles with a simple exchange of $s$ and
$t$. In other words the expression above includes the s--channel
contribution unlike field theory amplitudes where one needs to add
separately the other channels in order to derive a consistent
theory. In the tensionless case we have moved all massive poles of
the above expression to become massless. It is obvious that this
spoils the dual nature of string amplitudes. In order to obtain a
consistent theory in the tensionless limit it seems necessary to
include the collective contribution of the massive poles which
were pushed to the massless level.

We have pointed out before that the asymptotic behavior of string
amplitudes under BCFW deformations and their form in the Regge
regime are directly connected. So based on the above comments
regarding the Veneziano amplitudes it is instructive to look at
its Regge limit to see what might be the missing ingredient in
order to make the theory consistent under (\ref{BC1}). In string
theory we know that the Regge limit scattering is better described
using Pomeron operators which effectively average the
contributions of zeros and poles of a given amplitude as we take a
kinematic variable much larger than the others. This is a very
distinct behavior from the behavior of the amplitudes we studied
in section \ref{BCFWft}. Pomerons have a structure with a stringy
origin and amplitudes due to Pomeron exchanges present a diffusion
like behavior in transverse position space \cite{Brower:2006ea}.

Let us look first at the Pomeron states of string theory. They are
deduced using the OPE of normal string vertex operators
\cite{Brower:2006ea} and correspond to saddle point contributions
of string amplitudes in the Regge regime. For bosonic open strings
the OPE of two states with momenta $p_1$ and $p_2$ gives
\begin{equation}\label{pomeron}
{\cal V}_P \sim  C_n\ \Pi(\apr p^2) \ e^{ip\cdot X}[ q\cdot
\partial X]^{1- \apr p^2 }
\end{equation}
where $p=p_1+p_2$ and $q=p_1-p_2$.  The  propagator of the Pomeron is
\begin{equation}\label{pomprop}
\Pi(\apr p^2)=\Gamma\left(\apr p^2 -1\right)\ .
\end{equation}
and $C_n$ is an expression which depends on polarization and
momenta of the level $n$ states for which we consider the OPE.
Pomerons in string theory are physical states but with a
fractional oscillator number which renders them outside the normal
Hilbert space.
\begin{eqnarray} \label{Pomphys}
 L_0 {\cal V}_{P}&=& \apr p^2 + N-1=\apr   p^2 + \left(1-\apr  p^2\right)-1=0\ , \nonumber \\
 L_1 {\cal V}_{P}&\sim& q\cdot p= (p_1+p_2)\cdot (p_1-p_2)=0\ .
\end{eqnarray}
Notice that in the tensionless limit these states remain in the
spectrum.

The Veneziano scattering amplitude in the Regge regime $s \gg t$
can be computed either using the Pomeron vertex operator
(\ref{pomeron}) or directly from the Veneziano amplitude
\cite{Brower:2006ea, Cheung:2010vn}
\begin{eqnarray}\label{M4regge}
{\cal M}_4 &&\sim \int^{+\infty}_{-\infty} d y |y|^{-2+\apr t}
|1-y|^{-2+\apr s}
\sim \nonumber \\
&& \sim 2 \pi \ \Gamma(-1+\apr t)  (\apr s )^{1-\apr t}
\end{eqnarray}
Taking the tensionless limit $\apr \to \infty$ limit it implies
\begin{equation}\label{HSpomeron}
{\cal M}_4^t \sim {1\over \sqrt{t}}\left( {t\over
s}\right)^{\kappa^{-2} t-1} e^{-\kappa^{-2} t}
\end{equation}
where $\kappa$ the critical string length we have assumed in order
for our formulas to make sense. This should be compared to
(\ref{CEstregge}) where we see the marked difference of the two
expressions: for the region $s\gg t$ where the two expressions
apply one goes as $e^{-\kappa^{-2} s}$ and the other one has the
typical Pomeron behavior $s^{\kappa^{-2} t-1}$. Our discussion
suggests that massless current exchanges alone cannot lead into
the behavior above. We propose that the expression in
(\ref{HSpomeron}) is the appropriate asymptotic behavior, of the
tensionless limit of the four point string amplitude, for BCFW
deformations $s\sim-u\sim q z$.

Pomerons play role in the physics of scattering amplitudes in the
Regge region of kinematic variables, as in the example described
above. They appear as composite particles exchanged between
oppositely highly boosted external states. It is plausible
therefore that in the high energy limit we might have to consider
additional states which appear in the intermediate channels of
elementary particle scattering. Moreover it is possible that they
do not represent asymptotic states of the theory. Pomerons in QCD
represent coherent color-singlet objects built from gluons, which
are exchanged between hadrons in the Regge regime. Here we suggest
that there could be composite non-local objects in the theory of
massless HS fields which contribute to the scattering of HS fields
in the Regge regime.
 We emphasize that the true nature of
the advocated objects away from the Regge regime is not understood
at the moment. Moreover notice that the Pomeron in QCD like
theories emerges from a perturbative resummation of Feynman
diagrams while in the present context of the high energy limit of
string theory emerges at tree level in the HS theory. So in this
case they are not perturbation theory artifacts.

\section{Conclusions}

In the first part of this paper we demonstrated how one can build
off --shell cubic and quartic interaction vertices for the string
inspired systems which contain an infinite number of massless
bosonic Higher Spin fields. We hope that these results could be
useful for further investigation of consistency properties of
interacting Higher Spin fields on Minkowski background. In
particular, it would be interesting to perform explicit
calculations using perturbation theory and study the consistency
of higher order interactions.

Another interesting question is to consider an anti de Sitter
space--time background. In particular to find similar vertices on
an AdS background using the technique of \cite{Buchbinder:2006eq},
\cite{Fotopoulos:2007yq}, \cite{Fotopoulos:2006ci}
 and to discuss a possible connection with the results of \cite{Fradkin:1986qy}  and  implications
for AdS/CFT correspondence of these models.

In the second part of this paper we demonstrated how the BCFW
method for consistency checking of the S-matrix for a given theory
could shed some light to the construction of interacting Higher
Spin theories. The key point is that failure of the consistency
condition stated in \cite{Benincasa:2007xk}, for theories which
satisfy certain criteria under BCFW deformations, gives us some
information on the possible modification of the theory in order to
make it constructible. In essence it allows us to identify a
possible missing ingredient in order to make consistent Higher
Spin interactions in flat space--time. We showed that for two
candidate cubic couplings the BCFW consistency condition fails.

In one case, the {\it field theory} coupling, we argued that the
amplitude behaves well under BCFW deformation at infinite complex
momenta and therefore the BCFW relations should hold. Failure of
the condition (\ref{BC1}) was attributed to additional states
which might be needed in order to make the theory consistent with
BCFW. Moreover the analysis suggested that these states could be
extended objects.

In the second case, the {\it string theory} couplings, a naive
analysis based on Feynman diagrams for massless current exchanges
seems to imply that BCFW method should not be applicable.
Nevertheless recent progress in string theory suggests
\cite{Boels:2010bv, Cheung:2010vn} that BCFW method applies to
string theory amplitudes. There does not seem to be  a difficulty
in taking the tensionless limit of these analysis which in turn
implies that massless Higher Spin theory derived from the high
energy limit of string theory should also allow BCFW recursion
relations. Since the central object in the asymptotic behavior for
the tensile string analysis is the string Pomeron, we suggested
that it is plausible that one should supplement the theory with
non--local objects which have Pomeron like behavior in the Regge
regime of Higher Spin amplitudes.

The upshot of both cases is that interacting Higher Spin theories
in flat background with only point particle states although not
improbable, might not be the ones related to the high energy limit
of string theory. Although one can add perturbatively quartic and
higher vertices in the Lagrangian, consistently with gauge
invariance at each order, a non-trivial S-matrix, its analyticity
properties and BCFW constructibility seem to imply that: one
should consider Higher Spin theories in a larger frame which
includes extended and/or non-local objects in their spectrum.
\\

\noindent {\bf Acknowledgments.} We are grateful for numerous
discussions with X. Bekaert, P. Benincasa, F. Cachazo, A. Sagnotti
and M. Taronna. The work of A. F. was supported by an INFN
postdoctoral fellowship and partly supported by the Italian
MIUR-PRIN contract 20075ATT78. The work of M.T. has been supported
by a STFC rolling grant ST/G00062X/1.
\\

\noindent {\bf Note added.} When the present work was on its final
stage for submission, an updated version of the paper
\cite{Sagnotti:2010at} appeared in the archive which includes an
analysis having an overlap with ideas of sections \ref{Cubic} and
\ref{QUAR}. In our paper use the triplet formulation for reducible
Higher Spin fields in order to extend off--shell the  vertices
proposed in the original version of  \cite{Sagnotti:2010at}. On
the other hand in \cite{Sagnotti:2010at} they have used  the
compensator formulation , which describes irreducible Higher Spin
modes. The two results are closely related.


\renewcommand{\thesection}{A}

\setcounter{equation}{0}

\renewcommand{\theequation}{A.\arabic{equation}}

\end{document}